# Holstein polaron in a valley-degenerate two-dimensional semiconductor

Mingu Kang[1,2], Sung Won Jung[1,3], Woo Jong Shin[1,3], Yeongsup Sohn[1,3], Sae Hee Ryu[1,3], Timur K. Kim[4], Moritz Hoesch[4,5] & Keun Su Kim[1]★

[1]Department of Physics, Yonsei University, Seoul, Korea. [2]Department of Physics, Massachusetts Institute of Technology, Cambridge, MA, USA. [3]Department of Physics, Pohang University of Science and Technology, Pohang, Korea. [4]Diamond Light Source, Harwell Campus, Didcot, UK. Present address: [5]Deutsches Elektronen-Synchrotron (DESY), Notkestraße, Hamburg, Germany.

★e-mail: keunsukim@yonsei.ac.kr

**Two-dimensional (2D) crystals have emerged as a class of materials with tuneable carrier density[1]. Carrier doping to 2D semiconductors can be used to modulate manybody interactions[2] and to explore novel composite particles. Holstein polaron is a small composite particle of an electron carrying a cloud of self-induced lattice deformation (or phonons)[3–5], which has been proposed to play a key role in high-temperature superconductivity[6] and carrier mobility in devices[7]. Here, we report the discovery of Holstein polarons in a surface-doped layered semiconductor, $MoS_2$, where a puzzling 2D superconducting dome with the critical temperature of 12 K was found recently[8–11]. Using a high-resolution band mapping of charge carriers, we found strong band renormalizations collectively identified as a hitherto unobserved spectral function of Holstein polarons[12–18]. The unexpected short-range nature of electron-phonon (*e-ph*) coupling in $MoS_2$ can be explained by its valley degeneracy that enables strong intervalley coupling mediated by acoustic phonons. The coupling strength is found to gradually increase along the superconducting dome up to the intermediate regime, suggesting bipolaronic pairing in 2D superconductivity.**

A key to understanding the dynamics of polarons is their spectral function that has been studied with two standard models, Frölich polaron[4] and Holstein polaron[5]. The latter is a more realistic lattice model for polarons whose spatial extent can be either short-ranged within few atoms (small polaron with the *e-ph* coupling constant $\lambda > 1$) or long-ranged over several atoms (large polaron with $\lambda < 1$). Even this simple model is difficult to solve analytically, and the perturbation theory fails to account for the important intermediate regime ($\lambda \sim 1$) between small and large polarons. It was only recently that the spectral function of Holstein polaron at $\lambda \sim 1$ was successfully calculated in a nonperturbative manner by means of different theoretical techniques[12–18]. They have commonly predicted that electrons in a parabolic band (dotted lines in Fig. 1b) are scattered by the *n* number of phonons to form $e + n\text{-}ph$ continuums (orange and green regions). The state mixing with these continuums makes the parabolic band be separated into a series of sub-bands, which is accompanied by small gaps and flatting of dispersion at the integer multiple of phonon energy ($-n\Omega_0$, $n = 1, 2, 3, …$). The spectral weight is suppressed for the flat parts, so that the simulated spectral function (Fig. 1c) appears to be a series of broken bands, which is distinct from conventional kink[19], peak-dip-hump[19–21], and shake-off replica[22–25]. This characteristic spectral function of Holstein polarons should be observable with the use of angle-resolved photoemission spectroscopy (ARPES), but so far has been studied only in ultracold atomic gases[26] due to the lack of a suitable material system.

On the other hand, 2H transition-metal dichalcogenides (TMDs) have continued to attract broad interest as a class of two-dimensional (2D) van-der-Waals semiconductors with valley degeneracy[27,28]. This valley degree of freedom opens a channel of strong coupling between electronic states at two inequivalent valleys (K and K' in Fig. 1d) mediated by short-wavelength phonons (arrows)[27–30], which can potentially be used to study short-range e-ph coupling. Furthermore, electron doping to the surface of bulk (or multilayer) TMDs was found to develop the superconducting dome[8]. The critical temperature ($T_c$) as high as 12 K, which is a record among TMDs, was reported in $MoS_2$ (refs. 8–11), making it ideal to explore a signature of strong electron-boson coupling. There have been ARPES experiments on bulk or multilayer TMDs whose surface layer is chemically doped by the deposition of alkali-metals atoms[31–34], a well-known technique to form 2D electric dipole layers[34] as in ionic liquid gating[8–11] (Supplementary Fig. S1). However, no clear signature of electron-boson coupling has been reported so far. In this study, we carefully optimized samples and experimental conditions to measure ARPES data at high-resolution settings (see Methods for details), which can be used to detect a band-renormalization effect.

Figure 1d gives an overview of the low-energy band structure of surface-doped $MoS_2$, where ARPES intensity (shown in blue) is plotted as a function of energy and momentum ($k$). There is a well-defined valence band (VB) whose vertex is located at the zone centre ($k = 0$). Upon the deposition of Rb atoms, the donated electrons are known to be confined mostly within the topmost $MoS_2$ layer[34,35], leading to the formation of 2D electron gas, which is essentially the same situation where the 2D superconductivity was observed[8–11] (Supplementary Fig. S1). The band structure in the surface 2D layer of $MoS_2$ is shifted to the higher binding energy (Supplementary Fig. S2), and two minima of the conduction band (CB) appear below the Fermi energy ($E_F$) mainly at K and K' points and partly at T points, the latter of which can be more clearly identified in the Fermi surface map (upper panels). The CB minimum located at the T point for pristine bulk $MoS_2$ is moved to the K point due to the electric field effect[10], which is in agreement with the previous reports[33–35]. Below, we will focus on the K valleys of CBs, where e-ph coupling is found to be greater than at T points (Supplementary Fig. S3).

In a non-interacting picture, the local band structure of CBs at the K (or K') point can be approximated by a simple parabola with the effective mass $m_0 = 0.85 m_e$, where $m_e$ is the electron rest mass. This is shown by our spectral simulations in Fig. 1c for $\lambda = 0$ (left).

However, a striking deviation from this picture is observed in ARPES spectra (Fig. 1e) taken with high-resolution settings and low sample temperature (Methods). Instead of a simple parabolic band, we find strong distortions of the parabolic band at two energies (dotted lines), leading to three broken bands indicated by red, yellow, and green arrows. This unusual rearrangement of states is uniformly observed in different $k$ cuts about the K point (Supplementary Fig. S4), and not expected for structural and electrostatic origins, such as quantum confinement (Supplementary Fig. S5).

The spectral function of $MoS_2$ can be more closely examined with a series of energy-distributions curves (EDCs) shown in Fig. 2a. Whereas only a single pole is expected for each EDC in a non-interacting picture, even raw EDCs at $k_1$ and $k_2$ clearly show multiple poles separated by small gaps, a signature of band renormalizations by strong coupling to bosonic modes or polarons[12,13]. A curve fit to each EDC yields peak positions marked by closed circles whose size is scaled with their spectral weight in Fig. 2a,b. Around each small gap ($\delta_1$ and $\delta_2$ in Fig. 2b), the dispersion of poles is anti-crossing each other, and connected to flat branches with little spectral weight (Fig. 2c and Supplementary Fig. S6). A close pair of flat branches asymptotically approaches to either $\delta_1$ at $-26 \pm 3$ meV or $\delta_2$ at $-53 \pm 3$ meV, which are uniformly spaced at $-n\Omega_0$, where $\Omega_0 = 26$ meV. That is, the experimental spectral function consists of three characteristic sub-bands (shown in red, yellow, and green) as summarized in Fig. 2b.

In comparison to Fig. 1b, such a series of band renormalizations at $-n\Omega_0$ is the hitherto unobserved characteristic signature of Holstein polarons[12–18]. We employed a simple model of $k$-average approximation[17] to simulate the spectral function of Holstein polarons at $\lambda = 0.5$ (Fig. 1c and Supplementary Fig. S7), which successfully reproduces our experimental observations in Fig. 1e. This can be further corroborated by the same simulations that predict a series of flat dispersions at the onset of $e + n\text{-}ph$ continuums, albeit their weak spectral weight, to be observable, as shown in Fig. 2d. Such flat dispersions with a weak intensity are indeed observed in our experimental data taken for the corresponding region (Fig. 2e), which are marked by red, yellow, green, and blue arrows. The latter is more clearly seen in Fig. 2f as a shoulder structure at $-78 \pm 3$ meV or $-3\Omega_0$, and the energy of band renormalizations is confirmed again to be located at the integer multiple of $\Omega_0$. Consequently, our experimental data (Figs. 1e and 2e) collectively identify key aspects of the Holstein spectral function shown in Figs. 1b and 2b.

On the other hand, there is another flat feature near $E_F$ marked by the red arrow in Fig. 2e, which is not shown in Fig. 2d. A possible origin of this difference is a small gap opening at $E_F$, which is further supported by the subtle flattening of dispersion near $E_F$ in Fig. 2b and Supplementary Fig. S6. This gap may arise from the onset of superconductivity[8–11] or the correlation of many polarons[36], which are not included in our simulations in Fig. 2d.

A key clue to understand the mechanism of Holstein polarons in MoS$_2$ can be obtained from the involved boson mode. From its energy value $\Omega_0 = 26$ meV, we identify this mode as longitudinal acoustic (LA) phonons near zone corners (K and K'), where its phonon dispersion is locally flat like that of optical phonons (Supplementary Fig. S8). This short-wavelength branch of LA phonons has been proposed to mediate strong coupling between CBs at two valleys[27,29], as illustrated by arrows in Fig. 1b,d. Since the $k$ range of flat phonon dispersion is much wider than $\pm k_F$, this intervalley e-ph coupling can be reasonably approximated by the standard Holstein model. Therefore, the emergence of Holstein polarons in MoS$_2$ can be accounted for by short-range intervalley e-ph coupling as a natural consequence of its intrinsic valley degeneracy.

The important physical parameters of Holstein polarons can be quantified from the observed spectral function. A series of band renormalizations at $-n\Omega_0$ comes from mixing with an electron plus $n$-phonons continuums (Fig. 1b). The sub-bands whose energy is greater than $\Omega_0$ (yellow and green in Fig. 2b) are excited states of the resonant character, whereas that smaller than $\Omega_0$ corresponds to the polaron band (red). Along its dispersion, the spectral weight of poles in EDCs is plotted as a function of $k$ in Fig. 2c, which is proportional to quasiparticle residue $Z_k$. It abruptly drops from 1 at the dispersive to 0 at the flat branch, signalling a crossover from the e-like to the ph-like character within a single polaron band[12–18]. A polaron dragging a cloud of phonons (Fig. 1a) becomes heavier, and a fit to the polaron band with a simple quadratic function yields the mass enhancement factor as $m^*/m_0 \sim 3$. Comparing $m^*/m_0$ to theoretical simulations (those in Fig. 1c), we can roughly estimate $\lambda$ at $0.5 \pm 0.1$.

We now discuss the evolution of Holstein polarons with the doping density $\rho$. In Fig. 3a for low $\rho$, there is a narrow band with a broad tail toward the higher binging energy. Its EDC at $k_F$ in Fig. 3c consists of the main peak (in red) followed by a series of shake-off replicas, whose energy position is evenly spaced at $-n\Omega_0$ and whose intensity follows the Poisson distribution. This spectral function has been reported as a signature of long-range e-ph coupling or Frölich polarons in anatase TiO$_2$ (ref. 22) and SrTiO$_3$ (refs. 24,25). The long-range e-ph interaction is screened as the band width of filled states ($W$) exceeds $\Omega_0$ (ref. 37), that is, the inverse adiabaticity ratio $W/\Omega_0 > 1$. In contrast to conventional kinks within the framework of Migdal-Eliashberg theory[19,22,25], the valley degeneracy of MoS$_2$ enables a channel of strong short-range e-ph coupling, leading to the remarkably distinct signature of Holstein polarons (Fig. 3b). In this regime, the energy position of $\delta_1$ and $\delta_2$ remains nearly unchanged with $\rho$ at $-n\Omega_0$ and $\Omega_0 = 26 \pm 3$ meV (dotted lines in Fig. 3d). This shows the dominance of intervalley coupling to the lower-energy dispersionless part of acoustic phonons rather than to optical phonons, as predicted in theory[38].

On the other hand, the magnitude of $\delta_1$ and $\delta_2$ progressively increases with $\rho$ (Fig. 3d), suggesting the enhancement of short-range e-ph coupling. The corresponding $m^*$ of polarons (black lines) is plotted as a function of $\rho$ in Fig. 4a. While $m_0/m_e$ shows little dependence on $\rho$ in contrast to that of long-range *e-ph* coupling[25], $m^*/m_0$ increases by more than a factor of 5, much greater than theoretical predictions for many polarons at constant λ (ref. 39). This is consistent with the calculated intervalley *e-ph* coupling matrix that increases with *k* relative to the K point (dashed line in Fig. 4a)[29]. λ is estimated from $m^*/m_0$ to increase in the intermediate coupling regime (λ = 0.5 ~ 0.8), where the formation of bipolaron bound states is predicted[6]. Shown together in Fig. 4a is the superconducting dome in surface-doped $MoS_2$ (green area) reported using the ionic liquid gating[8–11]. Although its superconducting gap is too small (3–4 meV) to be resolved by our experiments, $\rho$ at which we find Holstein polarons of λ = 0.5 ~ 0.8 coincides with the onset of superconductivity.

These experimental findings suggest that intervalley coupling of polarons in exchange of acoustic phonons (bipolaronic coupling) may play a key role in the formation of Cooper pairs (Fig. 4b), which is compatible to the picture of intervalley singlet pairing[9,10]. Thus, our observation of Holstein polarons in $MoS_2$, along with the recent observations of its long-range counterparts in high-$T_c$ FeSe/$SrTiO_3$ (ref. 23) and $SrTiO_3$ (refs. 24,25), should be carefully considered in further discussions and theoretical studies on the mechanism of superconductivity[8–11]. The observation of Holstein polarons in $MoS_2$ may also be useful in understanding carrier mobility and dynamics in $MoS_2$-based devices (valleytronics)[27,28].

## Acknowledgements

This work was supported by the National Research Foundation (NRF) of Korea (Grants No. 2017R1A5A1014862, No. 2017R1A2B3011368), Future-leading Research Initiative of 2017-22-0059 of Yonsei University, and the POSCO Science Fellowship of POSCO TJ Park Foundation. This work was carried out with the support of the Diamond Light Source (beamline I05). The work at the Advanced Light Source was supported by the US Department of Energy, Office of Sciences under Contract No. DE-AC02-05CH11231. M.K. acknowledges the Samsung Scholarship from Samsung Foundation of Culture. We thank A. Bostwick, C. Jozwiak, and E. Rotenberg for help in ARPES experiments, and R.

Comin for helpful discussions.## Author Contributions

M.K. conducted experiments and analysed data with help from W.J.S., Y.S., S.H.R., T.K., and M.H. S.W.J. performed spectral-function simulations. K.S.K. directed the project. M.K. and K.S.K wrote the manuscript with contributions from all other co-authors.

## Competing interests

The authors declare no competing interests.

## Additional Information

**Supplementary information** is available for this paper at.

**Reprints and permissions information** is available at www.nature.com/reprints.

**Correspondence and requests for materials** should be addressed to K.S.K.

**Publisher's note:** Springer Nature remains neutral with regard to jurisdictional claims in published maps and institutional affiliations.

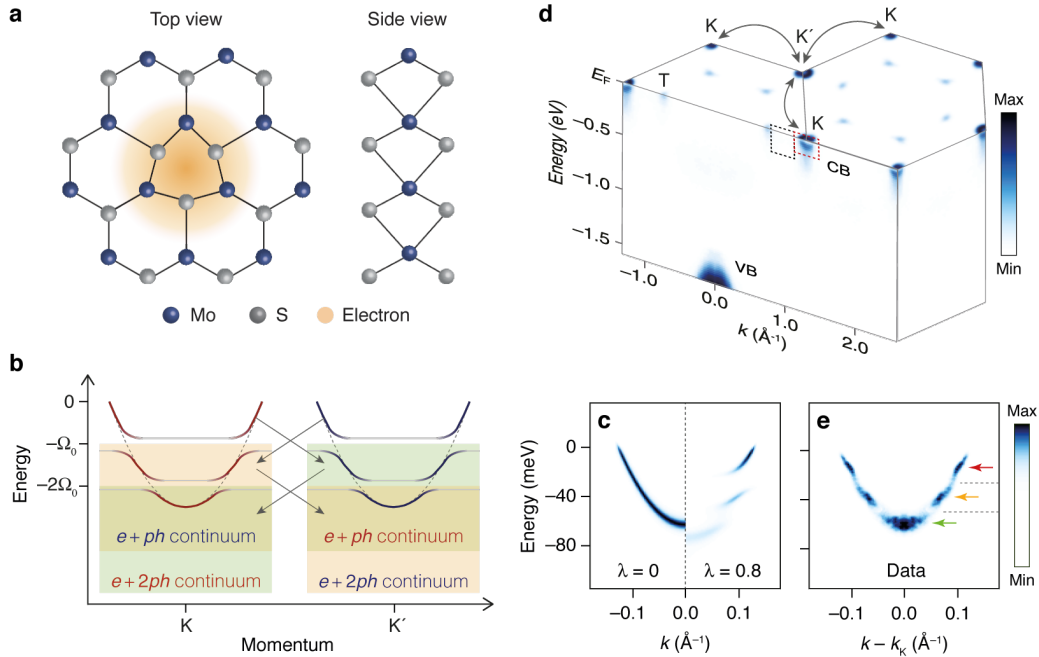

**Figure 1 | Holstein polaron of MoS$_2$. a**, Schematic illustration of a polaron, an electron carrying its surrounding lattice deformation, in MoS$_2$. **b,** Energy spectrum of Holstein polarons in MoS$_2$ resulting from strong coupling between K and K' valleys mediated by phonons (arrows). Dotted lines are the dispersion of bare bands in a non-interacting picture. Red and blue lines represent the electron character with $Z_k \sim 1$, whereas the thinner gray lines represent the phonon character with $Z_k \sim 0$. **c,** Simulated spectral function of Holstein polarons with zero ($\lambda = 0$) and intermediate ($\lambda = 0.5$) *e-ph* coupling based on *k*-average approximation (Methods)[17]. **d,** 3D representation of ARPES spectra taken for MoS$_2$ doped with Rb atoms at $\rho = 5 \times 10^{13}$ cm$^{-2}$. The photon energy was 54 eV, and the sample temperature was 80 K. **e,** High-resolution ARPES data taken at the region enclosed by the red dashed box in **d** and symmetrized with respect to the K point. Dotted lines indicate the energy position of gaps, while red, yellow, and green arrows indicate a series of sub-bands.

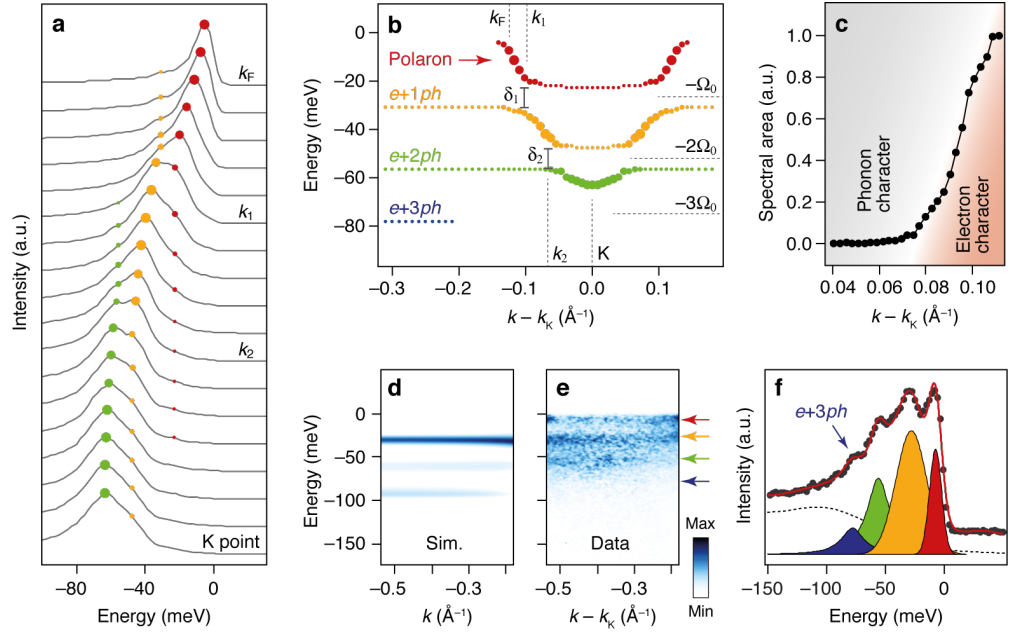

**Figure 2 | Spectral function of Holstein polarons. a,** Series of EDCs at $k$ ranging from the K point to the Fermi momentum ($k_F$) with 0.007 Å$^{-1}$ steps. Closed circles mark the peak positions of three sub-bands shown in Fig. 1e, and their size is roughly proportional to the spectral weight. **b,** Energy position of poles plotted as a function of $k$. $\delta_n$ is an anti-crossing gap at $-n\Omega_0$ and $k_n$. **c,** Spectral area of peaks plotted as a function of $k$ along the polaron dispersion shown in red in **b**, which is proportional to $Z_k$. **d,** Simulated spectral function at $\lambda = 0.5$ for the region of flat dispersions in **b** (Methods)[17]. **e,** High-resolution ARPES data taken for the region enclosed by the black dashed box between the T and K points in Fig. 1d, which is the same region as in **d**. Red, yellow, green, and blue arrows indicate the energy position of flat sub-bands shown in the corresponding colour in **b**. **f,** $k$-integrated EDC of data shown in **e**. Red line overlaid is the best fit with a background (dotted line) and four components shown in corresponding colours with the arrows in **e**.

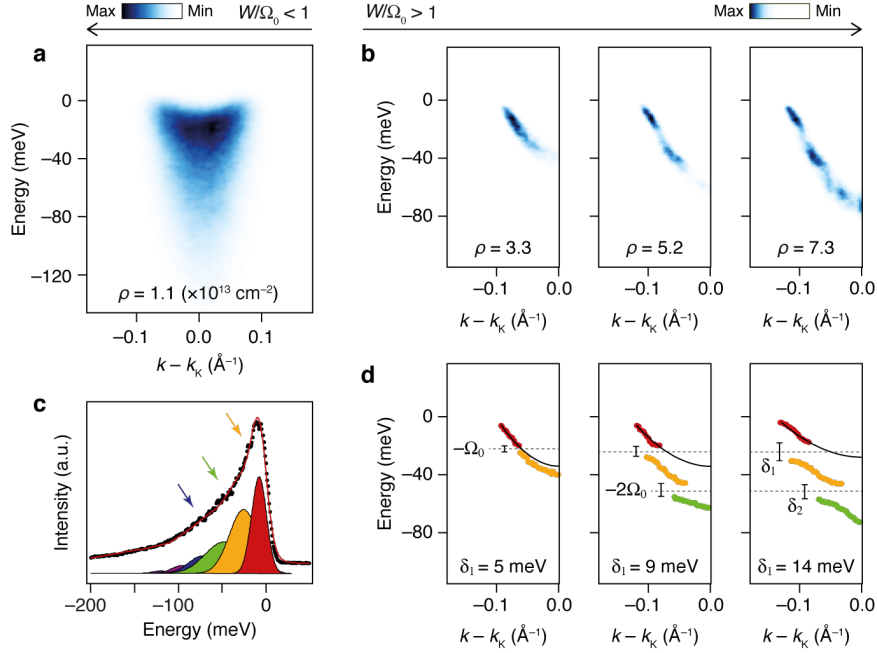

**Figure 3 | Doping dependence of polarons. a,b,** Series of ARPES spectra taken at $\rho$ marked at the bottom of each panel with the photon energy of 54 eV and the sample temperature of 7 K. The ratio of $W/\Omega_0$ separates two regimes, the unscreened (a) and the screened (b). **c,** EDC at $k_F$ from data shown in **a**. The red line overlaid is the best fit with a series of peaks at $-n\Omega_0$, whose intensity follows the Poisson distribution (Methods). **d,** Corresponding peak positions of those in **c**. Dotted lines indicate the energy positions of $-n\Omega_0$. Scale bars show the magnitude of $\delta_1$ and $\delta_2$, the former of which is shown at the bottom of each panel. Black curves overlaid are a fit using a simple quadratic function to the polaron dispersion (red dots) with negative and positive velocity, but only half of them with negative velocity is shown in **d**.

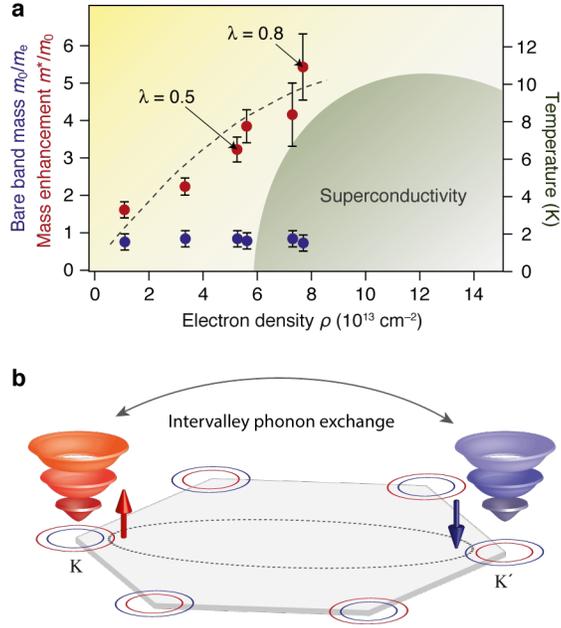

**Figure 4 | Strength of *e-ph* coupling and superconductivity. a,** Effective mass of polaron bands (closed circles) and bare bands (open circles) plotted as a function of $\rho$. The error bars are the range of $m^*/m_0$ obtained by curve fitting to data points in Fig. 3d with a simple quadratic function. The dashed line is the intervalley *e-ph* coupling matrix calculated for LA phonons near the K valleys of $MoS_2$ (ref. 29). The green region shows the superconducting dome of $MoS_2$ (ref. 9). **b,** Schematic illustration showing that two polarons with opposite spin at two K valleys are coupled via the exchange of phonons (bipolaronic coupling). Blue and red circles projected onto the Brillouin zone show the spin splitting in CB arising from spin-orbit coupling and interlayer symmetry breaking (Supplementary Fig. S2)[9,10]. The effect of this spin splitting to the spectral function of polarons is not considered here, since its energy scale is much lower than $\Omega_0$ (ref. 40).

# Methods

**Samples and chemical doping.** Single-crystal $MoS_2$ (99.995%, HQ Graphene) were glued by silver epoxy to sample holders, and cleaved in vacuum chambers with base pressure better than $4 \times 10^{-11}$ torr. The high crystal quality of samples was confirmed from the extremely narrow spectral width as shown in Supplementary Fig. S2. Surface doping was achieved by the *in situ* deposition of alkali metals on the surface of $MoS_2$ using commercial dispensers (SAES). To avoid intercalation, we chose Rb deposited on $MoS_2$ samples kept under 30 K (refs. 31–34). A Rb atom on $MoS_2$ donates charges mostly to the topmost layer, leading to the formation of 2D dipole layers or electric double layers[34,35]. This is essentially the same mechanism as in the ionic liquid gating (Supplementary Fig S1), which is also accompanied by interlayer symmetry breaking. The signature of interlayer symmetry breaking was clearly observed in Supplementary Fig. S2, ruling out the possibility of intercalation that would generate only the doping effect with no strong interlayer symmetry breaking. The electron density at the surface layer of $MoS_2$ is approximately estimated from the area enclosed by Fermi contours at $E_F$ based on Luttinger's theorem.

**ARPES experiments.** We carried out ARPES measurements at two synchrotron radiations, Beamline I05 in the Diamond Light Source and at Beamline 7.0.2 in the Advance Light Source. ARPES end-stations are equipped with R4000 hemispherical electron analysers. The photon energy was used in the range of 46 ~ 104 eV, and 54 eV was chosen for the maximal yield of photoelectrons. The light polarization was set to be either linear (horizontal or vertical) or circular, but we found essentially the same results, as shown in Supplementary Fig. S4. We have optimized samples, photon energy, and photoexcitation cross-section to achieve an unusually high yield of photoelectrons. This allows us to have enough photoelectrons to measure at high-resolution analyzer settings (more specifically, the analyzer slit of 0.1 mm). Owing to this level of energy resolution, the key signature of Holstein polaron (Fig. 1b) could be clearly observed even in raw data (Figs. 1e and 2e) even without taking any background subtraction or second derivatives.

**Spectral function analysis.** To estimate physical parameters of Holstein polarons, we performed a systematic and quantitative analysis to a series of EDCs ranging from the K point to $k_F$ (Fig. 2a and Supplementary Fig. S4). A fit to each EDC with multiple peaks (whose line shape is set to the Voight function, the use of either Lorentzian or Gaussian functions makes little difference) yields their energy positions (Fig. 2b) and spectral weight (Fig. 2c). The energy position of peaks is shown by closed circles, whose size is roughly proportional to their spectral weight in Fig. 2a,b. A fit to the bottom energy and $\pm k_F$ of CBs in Fig. 3d with a single quadratic function yields the effective mass of bare electrons $m_0/m_e$ (blue circles in Fig. 4a). On the other hand, the same fit to only those shown in red in Fig. 3d yields the mass enhancement of polarons (red circles in Fig. 4a). A fit to the $k$-integrated EDC (the red curve in Fig. 2f) with a series of peaks yields their energy positions at $-n\Omega_0$, where $\Omega_0 = 26 \pm 3$ meV. The relative intensity of satellite peaks ($n = 2, 3, \ldots$) to the main peak ($n = 1$) is additionally restricted to have the characteristic Poisson distribution as $a_c^{2n}/n!$ ($a_c$ = constant) in the fit to the EDC at $k_F$ in Fig. 3c.

**Spectral function simulations.** To calculate the spectral function of Holstein polarons, we employed an efficient analytical model called the *k*-average (MA) approximation[17]. This model is based on the standard Hamiltonian that describes an electron interacting with a dispersion-less phonon mode, that is, the Holstein polaron. For simplicity, we chose the 1D model, in which the *k*-averaged self-energy $\Sigma_{MA}$ over the Brillouin zone is simplified as

$$\Sigma_{MA}(\omega) = \cfrac{g^2 \Omega_0 \bar{g}_0(\omega - \Omega_0)}{1 - \cfrac{2g^2 \bar{g}_0(\omega - \Omega_0)\bar{g}_0(\omega - 2\Omega_0)}{1 - \cfrac{3g^2 \bar{g}_0(\omega - 2\Omega_0)\bar{g}_0(\omega - 3\Omega_0)}{1 - \cdots}}}$$

$$\bar{g}_0(\omega) = \frac{sgn(\omega)}{\sqrt{(\omega + i\eta)^2 - 4t^2}}$$

where $\omega$ is the energy, $\Omega_0$ is the photon energy taken from experimental data as 26 meV, $g^2 = 2t\Omega_0\lambda$, and $t$ is the hopping amplitude. This *k*-averaged $\Sigma_{MA}$ is taken to calculate the Green function as

$$G(k, \omega) = \frac{1}{\omega - E_k - \Sigma_{MA}(\omega) + i\eta}$$

where $E_k$ is the dispersion of the bare (undressed) band obtained from experimental data (Fig. 1e) as $m_0 = 0.85 m_e$ and $k_F = \pm 0.12$ Å$^{-1}$. By taking the imaginary part of $G(k, \omega)$, we calculated the single-particle spectral function $A(k, \omega)$ as a function of $\lambda$. $\lambda$ is set to 0 for the non-interacting case and 0.5 ~ 1.0 for the intermediate coupling regime.

**Data availability.** The data that supports the findings of this study are available from the corresponding author on reasonable request.